# STAGER checklist: Standardized Testing and Assessment Guidelines for Evaluating Generative AI Reliability

Jinghong Chen[1,*], Lingxuan Zhu[1,2,*], Weiming Mou[1,3,*], Zaoqu Liu[4,5], Quan Cheng[6,7,#], Anqi Lin[1,#], Jian Zhang[1,#], Peng Luo[1,#]

1 Department of Oncology, Zhujiang Hospital, Southern Medical University, 253 Industrial Avenue, Guangzhou, 510282, Guangdong
2 Department of Etiology and Carcinogenesis, National Cancer Center/ National Clinical Research Center for Cancer/Cancer Hospital, Changping laboratory, Chinese Academy of Medical Sciences and Peking Union Medical College, Beijing, China.
3 Department of Urology, Shanghai General Hospital, Shanghai Jiao Tong University School of Medicine, Shanghai, China.
4 State Key Laboratory of Proteomics, Beijing Proteome Research Center, National Center for Protein Sciences (Beijing), Beijing Institute of Lifeomics, Beijing 102206, China
5 State Key Laboratory of Medical Molecular Biology, Institute of Basic Medical Sciences, Chinese Academy of Medical Sciences, Department of Pathophysiology, Peking Union Medical College, Beijing, 100730, China
6 Xiangya Hospital, Central South University, Changsha, Hunan, China
7 National Clinical Research Center for Geriatric Disorders, Xiangya Hospital, Central South University, Changsha, China

[*]These authors contributed equally to this work

[#]Corresponding author:
Dr. Quan Cheng
Xiangya Hospital, Central South University, Changsha, Hunan, China
National Clinical Research Center for Geriatric Disorders, Xiangya Hospital, Central South University, Changsha, China
E-mail address: chengquan@csu.edu.cn
Dr. Anqi Lin
Department of Oncology, Zhujiang Hospital, Southern Medical University, 253 Industrial Avenue, Guangzhou, 510282, Guangdong
E-mail address: smulinanqi0206@i.smu.edu.cn
Dr. Jian Zhang
Department of Oncology, Zhujiang Hospital, Southern Medical University, 253 Industrial

Avenue, Guangzhou, 510282, Guangdong

E-mail address: zhangjian@i.smu.edu.cn

Dr. Peng Luo

Department of Oncology, Zhujiang Hospital, Southern Medical University, 253 Industrial Avenue, Guangzhou, 510282, Guangdong

E-mail address: luopeng@smu.edu.cn

# STAGER checklist: Standardized Testing and Assessment Guidelines for Evaluating Generative AI Reliability

## Abstract

Generative Artificial Intelligence (AI) holds immense potential in medical applications. Numerous studies have explored the efficacy of various generative AI models within healthcare contexts, but there is a lack of a comprehensive and systematic evaluation framework. Given that some studies evaluating the ability of generative AI for medical applications have deficiencies in their methodological design, standardized guidelines for their evaluation are also currently lacking. In response, our objective is to devise standardized assessment guidelines tailored for evaluating the performance of generative AI systems in medical contexts. To this end, we conducted a thorough literature review using the PubMed and Google Scholar databases, focusing on research that tests generative AI capabilities in medicine. Our multidisciplinary team, comprising experts in life sciences, clinical medicine, medical engineering, and generative AI users, conducted several discussion sessions and developed a checklist of 23 items. The checklist is designed to encompass the critical evaluation aspects of generative AI in medical applications comprehensively. This checklist, and the broader assessment framework it anchors, address several key dimensions, including question collection, querying methodologies, and assessment techniques. We aim to provide a holistic evaluation of AI systems. The checklist delineates a clear pathway from question gathering to result assessment, offering researchers guidance through potential challenges and pitfalls. Our framework furnishes a standardized, systematic approach for research involving the testing of generative AI's applicability in medicine. It enhances the quality of research reporting and aids in the evolution of generative AI in medicine and life sciences.

## INTRODUCTION

Generative AI, an increasingly prominent subfield of artificial intelligence[1], boasts the remarkable ability to generate data across diverse formats, including text, images, audio, video, and code[2]. This versatility extends to its real-time adaptability to novel task requirements through straightforward textual prompts[3]. In the realm of medicine, Generative AI stands out for its proficiency in rapidly processing multimodal information, such as medical texts and images. It can then deliver responses to medical inquiries in natural language, offering critical support to medical professionals in diagnostic decision-making and scientific research.

Current research on the application of generative AI in the medical field encompasses a broad spectrum, ranging from assessing its grasp of medical knowledge and ability to pass medical examinations[4], to aiding in providing initial medical counseling[5] and swiftly providing pertinent medical information and recommendations[6,7]. These studies underscore the vast potential for generative AI's deployment in healthcare. Nonetheless, a notable concern is that some published studies might exhibit methodological shortcomings and limitations in their assessment approaches. This can introduce varying degrees of bias into their findings. For instance, Fijačko et al. investigated ChatGPT's performance on the American Heart Association (AHA) Basic Life Support (BLS) and Advanced Cardiovascular Life Support (ACLS) exams and highlighted that ChatGPT was unable to pass these tests[8]. This conclusion, however, overlooked the fact that generative AI models often yield different responses to identical queries. A revised approach, involving the repetition of the same question, revealed that ChatGPT could indeed pass both exams with notable success[9]. Another critical gap lies in the lack of established frameworks for the systematic evaluation of generative AI in its capacity to address and apply solutions to medically relevant problems. Therefore, the formulation of comprehensive evaluation guidelines is crucial. Such a framework would not only standardize assessments but also significantly advance research in the realm of generative AI applications in medicine.

In our guidelines, we establish a standardized methodological framework designed to evaluate the applicability of generative AI systems in medical-related fields. This framework serves as a comprehensive guide for the medical assessment of generative AI technologies, including gathering questions, framing them appropriately, conducting thorough outcome assessments, and so on. Recognizing the variation in generative AI's performance between multiple-choice and open-ended questions, our guide thoughtfully differentiates the approaches for handling these two question types. This distinction ensures a more nuanced and effective evaluation process. Covering critical aspects of the research process, our guidelines aim to assist researchers, medical professionals, and technology developers in conducting a thorough and precise evaluation of generative AI's capabilities in medical aptitude assessments, which includes scrutinizing aspects such as accuracy, integrity, and readability.

## METHODS

To formulate assessment guidelines for evaluating generative AI's proficiency in medical competency testing, we initiated our research by conducting a comprehensive search for relevant studies in this area, utilizing the PubMed and Google Scholar databases. This exploration led to the extraction of a series of potential checklist items, meticulously designed to comprehensively cover the crucial assessment dimensions of generative AI for competency-related testing in medical applications. To validate and refine these checklist items, our interdisciplinary team—comprising specialists in life sciences, clinical medicine, and medical engineering, who are also generative AI users—engaged in a series of in-depth discussions. These deliberations were guided by the established protocols in the "Guidance for Developers of Health Research Reporting Guidelines"[10], ensuring a professional and pertinent approach. Throughout multiple sessions, the team meticulously reviewed, debated, and fine-tuned these prospective checklist items, to develop a cohesive and actionable assessment framework. In these meetings, members concentrated on scrutinizing each item on the checklist, contributing insights and suggestions. This rigorous and collaborative process guaranteed that every checklist item was thoroughly examined and critically assessed, ensuring a comprehensive and robust set of guidelines for evaluating generative AI in medical competency.

# RESULTS

We have formulated a checklist comprising 23 items (Table 1), which lays out an extensive framework for evaluating the proficiency of generative AI in medical contexts. This framework spans various crucial dimensions, including question collection, the approach to questioning[11], and diverse assessment methods (encompassing accuracy integrity, and readability). Such a comprehensive scope ensures a thorough assessment of generative AI's capabilities in handling medical data and scenarios. Delving into these dimensions has granted us a deeper insight into the strengths and potential limitations of AI in processing and interpreting medical information. Additionally, we offer detailed explanations of each item on the checklist (Table 2), elucidating the rationale behind every step. This guidance is designed to assist researchers in navigating the multifaceted challenges they might encounter throughout their investigative endeavors.

# DISCUSSION

These guidelines are centered around the creation of a meticulously developed 23-item checklist, tailored to evaluate generative AI's applicability in medicine and life sciences. The innovation of this guide lies in its provision of broad assessment dimensions for generative AI applications within these fields. It encompasses key aspects such as question collection, questioning approaches, and diverse assessment methods, which is holistic, facilitating a deeper understanding and assessment of generative AI's performance in medical contexts, thereby advancing the field.

The checklist was crafted with a keen insight into the current challenges and issues in applying generative AI within the medical field. With the rapid development of AI technology in medicine, its potential for answering medical queries has become increasingly apparent. Yet, significant challenges persist in this area's research and application, such as the opacity of generative AI's data processing and information generation[12], which can compromise the decipherability and interpretability of outcomes. This checklist offers a standardized framework to assess these critical issues, thereby enhancing the quality and reliability of relevant research. Additionally, a detailed explanation of the checklist in our guidelines not only aids researchers in comprehending the assessment process, minimizes subjective interpretation variances, and bolsters the reproducibility of the entire assessment process, but also empowers them to identify and tackle potential challenges during their studies, which is crucial for elevating research quality.

# CONCLUSIONS

The assessment framework outlined in these guidelines offers a standardized and systematic approach for evaluating research on generative AI in medical applications, aiming to enhance the quality of research reports. This framework is instrumental in fostering the development of generative AI within medical contexts and ensuring the validity and reliability of AI systems in practical use. It is anticipated that this framework will encourage academic collaboration and exchange, thereby contributing to the sustained advancement of generative AI technology in medical applications.

# Table 1. Evaluations of generative AI for medical applications.

| Section/Topic | Item No | Recommendation | Reported on Page No |
|---|---|---|---|
| **Title and abstract** | 1 | Identify the purpose of the research, the generative AI model used and its version, the source of the questions, methods, results and conclusions. | |
| **Introduction** | | | |
| Background | 2 | Explain the background and purpose of the study. | |
| Objectives | 3 | State specific objectives, including generative AI model used and its version, the training set used for generative AI, the source of the questions, the nature of research and the limitation. | |
| **Methods** | | | |
| Question collection | 4 | Select the professional questions from guidelines, medical examination question banks, high-frequency issues found via search engines like Google, or drafted by experts, ensure that the questions cover specific subfields of medicine. | |
| | 5 | Make sure the questions are representative in terms of difficulty, type and professionalism. | |
| | 6 | Describe how the questions were collected, the number of questions, whether the questions were pre-screened, the conditions of the screening, the modality of the input as well as the relevant format. | |
| Questioning | 7 | Use a consistent prompt with identically formatted patterns and provide the full prompt in the article. | |
| | 8 | Ask the same question multiple times, and record each response. | |
| | 9 | Indicate whether the question is open-ended or multiple choice. | |
| | 10 | Initiate a new chat for each question. | |
| | 11 | Record the data the responses were collected and the version of the generative AI. | |
| Accuracy | 12 | Describe any methods employed for scoring accuracy, such as employing a Likert scale when dealing with subjective questions. | |
| | 13 | Compare with reference answer, record the number of correct answers to each question and calculate correct rate if you asked objective questions. | |
| Integrity | 14 | Describe any methods used to access the integrity, such as using a Likert scale. | |
| Readability | 15 | Describe any methods used to access the readability, such as using the Flesch-Kincaid Readability Tests. | |
| Consistency | 16 | Assess consistency and reliability of reviewer ratings, avoiding significant differences in the subjective scores among reviewers. | |
| | 17 | Evaluate consistency across responses to the same question to assess whether the generative AI can steadily provide consistent response. | |
| **Results** | | | |

| | | |
|---|---|---|
| Results | 18 | Describe results for accuracy, completeness, and readability, recommending the use of tables or charts for presentation. |
| **Discussion** | | |
| Analyse | 19 | Analyse the results according to the study objectives. |
| Limitations | 20 | Explore the constraints of the research, acknowledging possible origins of partiality or inaccuracy. |
| | 21 | Engage in rational discussion and reject exaggeration. |
| Conclusion | 22 | Provide a condensed conclusion that summarizes the study's main findings, reiterates its importance, and indicate directions or recommendations for future research. |
| **Other Information** | | |
| Funding and sponsorship | 23 | Provide the origin of financial support and the function of the sponsors for the current investigation, as well as for the initial research if relevant to the foundation of this article. |

## Table 2. Explanations for evaluations of generative AI for medical applications.

| Section/Topic | Item No | Recommendation | Explanations |
|---|---|---|---|
| **Title and abstract** | 1 | Identify the purpose of the research, the generative AI model used and its version, the source of the questions, methods, results and conclusions. | Lay the foundation for readers to quickly understand the study and facilitate other researchers to critically analyze the design and results of this research. |
| **Introduction** | | | |
| Background | 2 | Explain the background and purpose of the study. | Enable readers to grasp the central theme of the article. |
| Objectives | 3 | State specific objectives, including generative AI model used and its version, the training set used for generative AI, the source of the questions, the nature of research and the limitation. | Provide the necessary framework for readers to understand the article. |
| **Methods** | | | |
| Question collection | 4 | Select the professional questions from guidelines, medical examination question banks, high-frequency issues found via search engines like Google, or drafted by experts, ensure that the questions cover specific subfields of medicine. | For guidelines or question banks, questions can be either manually selected or extracted using software, while using an API to select questions can reduce subjective errors and make more sense for the entire dataset. When selecting questions from search engines, researchers may opt for frequently occurring ones. If the questions is drafted by experts, the experts need to have authority and experience in the relevant field. |
| | 5 | Make sure the questions are representative in terms of difficulty, type and professionalism. | Enhance the universality of the study. |
| | 6 | Describe how the questions were collected, the number of questions, whether the questions were pre-screened, the conditions of the screening, the modality of the input as well as the relevant format. | Input modes such as text, image, sound, video input, etc., and related attributes (e.g. image resolution). |
| Questioning | 7 | Use a consistent prompt with identically formatted patterns and provide the full prompt in the article. | Reduce the objective differences introduced by different questioning methods and the impact of such differences on the quality of answers. Providing the full prompt in the article ensures that the study is transparent and reproducible. |
| | 8 | Ask the same question multiple times, and record each response. | Generative AI, known for delivering varied responses to identical queries, necessitates repeated questioning to gauge its consistency. |
| | 9 | Indicate whether the question is open-ended or multiple choice. | Subjective and objective questions are assessed differently. |
| | 10 | Initiate a new chat for each question. | Prevent generative AI from being affected by context. |
| | 11 | Record the data the responses were collected and the version of the generative AI. | Reduce the impact of performance differences between AI versions and the timing of knowledge updates. |
| Accuracy | 12 | Describe any methods employed for scoring accuracy, such as employing a Likert scale when dealing with subjective questions. | Accuracy refers to the degree to which the response reflects or corresponds to reality or truth. |
| | 13 | Compare with reference answer, record the number of correct answers to each question and calculate correct rate if you asked objective questions. | The more times the generative AI model response correctly, the more robust it is considered to be. |
| Integrity | 14 | Describe any methods used to access the integrity, such as using a Likert scale. | Integrity refers to whether the response is comprehensive, detailed and covers relevant information. |
| Readability | 15 | Describe any methods used to access the readability, such as using the Flesch-Kincaid Readability Tests. | Reflect the ease with which a text can be read and understood (e.g. clarity of language, the organization of structure, grammatical and spelling accuracy and coherence). |
| Consistency | 16 | Assess consistency and reliability of reviewer ratings, avoiding significant differences in the subjective scores among reviewers. | Ensure the fairness and effectiveness of the evaluation process. |
| | 17 | Evaluate consistency across responses to the same question to assess whether the generative AI can steadily provide consistent response. | A way to effectively monitor model performance, helping detect if the model has erratic behavior. |
| **Results** | | | |
| Results | 18 | Describe results for accuracy, completeness, and readability, recommending the use of tables or charts for presentation. | Uncover its performance in specific subdomains is critical to a deeper understanding of the value and limitations of AI applications in medicine. |
| **Discussion** | | | |
| Analyse | 19 | Analyse the results according to the study objectives. | Comprehensively analyze the performance of generative AI in terms of accuracy, completeness, and readability. |
| Limitations | 20 | Explore the constraints of the research, acknowledging possible origins of partiality or inaccuracy. | Enhance the understanding of the scope, accuracy, and applicability of the research findings |
| | 21 | Engage in rational discussion and reject exaggeration. | Honest and rational expression is necessary to maintaining academic norms and advancing knowledge. |
| Conclusion | 22 | Provide a condensed conclusion that summarizes the study's main findings, reiterates its importance, and indicate directions or recommendations for future research. | Provide direction for future research and help promote the further development and application of generative AI in the medical field. |
| **Other Information** | | | |
| Funding and sponsorship | 23 | Provide the origin of financial support and the function of the sponsors for the current investigation, as well as for the initial research if relevant to the foundation of this article. | Maintain the objectivity and transparency of the research. |